\pdfoutput=1
\documentclass{llncs}

\usepackage{multirow}
\usepackage{algorithm}
\usepackage{algorithmic}
\usepackage{graphicx}
\usepackage{float}

\begin{document}

\title{Exploiting Redundancy, Recurrency and Parallelism: How to Link Millions of Addresses
with Ten Lines of Code in Ten Minutes}
\titlerunning{Hamiltonian Mechanics}  
%
\author{Yuhang Zhang \and Tania Churchill \and Kee Siong Ng}
\authorrunning{Yuhang Zhang et al.} 
%
%
\institute{Australian Transaction Reports and Analysis Centre\\
\email{yuhang.zhang@austrac.gov.au},\\ 
Canberra, Australia}

\maketitle              

\author{Yuhang Zhang
\and
Tania Churchill
\and
Kee Siong Ng}
%

\begin{abstract}
Accurate and efficient record linkage is an open challenge of particular
relevance to Australian Government Agencies, who recognise that so-called
wicked social problems are best tackled by forming partnerships founded on
large-scale data fusion.Names and addresses are the most common attributes on
which data from different government agencies can be linked. In this paper, we
focus on the problem of address linking. Linkage is particularly problematic
when the data has significant quality issues. The most common approach for
dealing with quality issues is to standardise raw data prior to linking. If a
mistake is made in standardisation, however, it is usually impossible to
recover from it to perform linkage correctly. This paper proposes a novel
algorithm for address linking that is particularly practical for linking large
disparate sets of addresses, being highly scalable, robust to data quality
issues and simple to implement. It obviates the need for labour intensive and
problematic address standardisation. Empirical results show that approximately
$91\%$  of the generated links created by matching two large address datasets
from two government agencies, were correct. Finally, we demonstrate that the
linking can be performed in under 10 minutes, with 10 lines of code.
\end{abstract}
\vspace{.1in}


\section{Introduction}

Efficient record linkage is an important step in large-scale automated data fusion. Data fusion is  a problem of increasing significance in the context of Australia's whole-of-government approach to tackling our most pressing social issues - including terrorism and welfare fraud - by combining and analysing datasets from multiple government agencies.
Outside of personal identifiers like tax-file numbers and driver's licenses, names and addresses are the two most important attributes on which disparate datasets are matched.
Whereas the problem of linking names is well-studied and there are specialised similarity measures like Jaro-Winkler for names \cite{stringmetrics}, not a great deal is known in the literature \cite{Christen:2012:DMC:2344108} about best practices for matching addresses, especially address data with significant quality issues.
%
Listed here are some address-specific challenges for efficient record linkage:
\begin{itemize}\itemsep1.5mm\parskip0mm
  \item \textbf{Incompleteness}: incomplete addresses
  that have no street type, no suburb name, no postcode, etc are common in address data.

\item \textbf{Inconsistent Formats}: the structure of addresses can be different between countries, regions and languages. 
  People can use variants to denote the same address, \textit{e.g.}, unit 1 of 2 Elizabeth Street, 1/2
  Elizabeth St, and U-1 2 Elisabeth Str all denote the same address.
  The presence of foreign characters in international addresses can also introduce issues.
  
  \item \textbf{Errors}: Wrong street types, invalid postcodes, non-matching suburb-postcode pairs, and
    various misspellings are widely seen in address data.

  \item \textbf{Unsegmented Addresses} Depending on the source, addresses can be captured as a single line of text with no explicit structural information.
    
\end{itemize}
These data quality issues may make two equivalent addresses look different and,
by chance, make two different addresses look similar.

The most common way to tackle data quality issues is to standardise raw
addresses before the linking operation \cite{Christen:2012:DMC:2344108}.
Address standardisation usually includes two types of operations: parsing and transforming.
With the parsing operation, addresses are parsed into semantic components, such as street, suburb, state, and country. For example, if an address contains three
numbers they are in order unit number, street number, and postcode.
In the transforming operation, variants of the same entity are transformed to a
canonical format and typos are removed, \textit{e.g.},
transforming Street, St, and Str all to Street.

The issue with standardisation is that it is in itself a challenging
problem. For example, \textit{Service Centre St George} might be interpreted
as a business name \textit{Service Centre of Saint George}; or a street name and a suburb
name \textit{Service Centre Street, George}; or a different business name
\textit{Service Centre of Street George}; or a suburb name and a state name
\textit{Service Centre, Saint George}.
Three numbers in an address can also be street number, level number in a high-rise, and postcode.
Intepreting an address is by nature ambiguous.

Address standardisation can be done using a rule-based system, or it can be done using machine learning approaches like Hidden Markov Models \cite{Christen2004,Christen05automatedprobabilistic}.
Ongoing research is still being undertaken to
improve standardisation accuracy~\cite{Guo:2009:ASL:1557019.1557144}.
Perhaps the biggest drawback of address standardisation is that if a mistake is made during standardisation, it is usually hard to recover from it to perform linkage correctly. Rule-based standardisation also tends to be specific to the individual dataset, failing to generalise well. 

\subsection*{Using Redundancy to Avoid Standardisation}
\noindent 
Instead of standardising raw addresses
into canonical forms, we rely on the redundancy in addresses to resolve data quality issues. 

We say an address contains redundancy if an incomplete
representation is sufficient to uniquely identify this address. For example, if
there is only one building in Elizabeth St that has Unit 123, then \textit{U123
45 Elizabeth St} as an address contains redundancy, because specifying street
number 45 is not really necessary. Redundancy exists widely in addresses. Not
every suburb is covered by postcode 2600. Not every state has a street named
Elizabeth. As an extreme example, three numbers like 18 19 5600, might be
enough to identify a unique address globally, as long as no other addresses
contain these three numbers simultaneously. Note that in this case, we do not
even need to know whether 18 is a unit number or a street number.

Our working hypothesis is that address data, in general, contains enough redundancy such that:
\begin{enumerate}\itemsep1mm\parskip0mm
  \item each address is still unique even when meta-data distinguishing address components such
  as street, suburb, and state are missing.
  \item equivalent addresses are
  still more similar to each other than to irrelevant addresses in the presence of
  errors or variants.
\end{enumerate}
Our assumptions - which stem from earlier experiments using compressed sensing techniques \cite{CPA:CPA20124} to represent and link addresses - are really stating that despite the data quality issues in
addresses, two addresses, in their raw form, can still be separated/linked if they are
different/equivalent.
In particular, address segmentation - a problem that is arguably as difficult as the general address-linking problem - and address standardisation are not strictly necessary.



\subsection*{Using Recurrency for Data-Driven Blocking}
When linking two large databases, algorithm efficiency is as important as
algorithm accuracy. An algorithm that takes days to finish
is not only too expensive to deploy, but is also infeasible to repetitively
evaluate during development. 

Blocking is a widely used technique to improve linkage efficiency.
Na{\"i}vely, linking two databases containing $m$ and $n$ addresses respectively requires
$O(mn)$ comparisons. Most of these comparisons lead to non-matches. To reject
these non-matches with a lower cost, one may first partition the raw addresses
according to criteria selected by a user. These criteria are called blocking
keys, which may be postcode, suburb name, \textit{etc}.. During linkage,
comparison is only carried out between addresses that fall into the same
partitions, based on the assumption that addresses which don't share a blocking key are not a match.

Blocking key selection largely determines the efficiency and completeness of
address linkage. If the keys are not meaningful, they will not help find matches and may even 
slow down the matching process. If too few keys are used, effeciencies won't be gained.
If too many keys are used, one may fail to discover all possible links.
If different blocking keys do not distribute evenly among the addresses,
the largest few partitions will form the bottleneck of linkage efficiency.
Moreover, the performance of blocking keys in previous work also depends on
the accuracy of address standardisation.

In the spirit of \cite{halevy09}, we propose in this paper a data-driven approach to select blocking keys based on their
recurrency. These data-driven blocking keys are by design adapted to the
database at hand, statistically meaningful as address differentiators, evenly distributed,
and provide comprehensive cover to all addresses.
Since we implement no standardisation, our blocking keys do not depend on the success of standardisation either. 

\subsection*{Implementation on Parallel Platforms}
Massively parallel processing databases like Teradata and Greenplum have long supported parallelised SQL that scales to large datasets.
Recent advances in large-scale in-database analytics platforms \cite{DBLP:journals/pvldb/HellersteinRSWFGNWFLK12}, \cite{Zaharia:2010:SCC:1863103.1863113} have shown us how sophisticated machine learning algorithms can be implemented on top of a declarative language like SQL or MapReduce to scale to petabyte-sized datasets on cluster computing.
Building on the same general principle, we propose in this paper a modified inverted index data structure for address linking that can be implemented in less than ten SQL statements and which enjoys tremendous scalability and code maintability.

\subsection*{Paper Contributions}

The paper's contribution is a novel address-linkage algorithm that:
%
\begin{enumerate}
  \item links addresses as free-text (including international addresses), obviating the need for labour-intensive and sometimes problematic address standardisation; 

  \item uses data-driven blocking keys to minimise unnecessary pairwise comparisons, in a way that obviates the need for address segmentation and avoids the usual worst-case scenarios encountered by using a fixed blocking key like suburb or postcode;

  \item introduces an extension of the inverted index data structure that allows two large address datasets to be linked efficiently; 
    
  \item is practical because of its simplicity, allowing the whole algorithm to be written in less than 10 standard SQL statements; and

  \item is scalable when the SQL statements are implemented on top of parallel platforms like the Greenplum Database (open-source parallel PostgreSQL) and Spark.
\end{enumerate} 
The algorithm is particularly suitable for integrating large sets of disparate address datasets with minimal manual human intervention.
It is also possible to combine the algorithm with a rule-based system to produce a model-averaging system that is more robust than each system in isolation.\\

The remaining sections of this paper are organised as follows.
We first explain how we link a single address to an address
database utilising redundancy. We then show how the same algorithm can be
carried out in batch taking advantage of recurring address components. We then
demonstrate the performance of our algorithm with two address linkage
applications, followed by our conclusion.

\section{Address as Bag of Tokens}
Without subfield structures, an address becomes a bag (or a multiset) of unordered tokens.
For example, 
\begin{table}[H]
\centering\small
\begin{tabular}{l|l|l|l|l}
\hline
No.&street&suburb&state&postcode\\
\hline
513& Elizabeth St& Melbourne& VIC & 3000\\ 
\hline
\end{tabular}
\end{table}
becomes 
\begin{eqnarray*}
\{\mbox{ 3000},\mbox{ 513},\mbox{ Elizabeth},\mbox{
Melbourne},\mbox{ Street},\mbox{ VIC}\}~,
\end{eqnarray*}
In this example, we implicitly define a token to be a word, or a maximal
character sequence that contains only letters and numerics. We can also define
a token to be a single character, 
\begin{eqnarray*}
\{0,0,0,1,3,3,5,\mbox{a,b,b,c,e,e,e,e,}
\mbox{h,i,l,l,n,o,r,r,s,t,t,t,u,v,z}\}~,
\end{eqnarray*}
a two-word phrase,
\begin{eqnarray*}
\{\mbox{ 513 Elizabeth},\mbox{ Elizabeth St},
\mbox{ St Melbourne},\mbox{Melbourne VIC},\mbox{ VIC 3000}\}~.
\end{eqnarray*}
or generally anything we like. Note that in the above example, two-word
phrases preserve pairwise order information in the original address. We can
also use two word tokens that do not contain pairwise order information.

Different types of tokens have different distinctiveness powers and different
tolerances against data quality issues. To see the difference, note the word
token, `Melbourne', can match to any appearance of `Melbourne' in other
addresses, such as Melbourne Avenue, Mount Melbourne, Melbourne in Canada, \textit{etc.}. By contrast, the phrase token,
`Melbourne VIC', can only match the co-occurence of `Melbourne' and `VIC'.
The advantage of being distinctive is that we can reduce false matches. The
disadvantage, however, is that we may miss a true match if the other address did not include the state information of `VIC'
or included it in a different form, \textit{e.g.}, Victoria. 

For the purposes of linkage, we do not need individual tokens to be distinctive.
Instead, we want tokens to be tolerant to data quality issues. We lose nothing
as long as a bag of tokens as a whole is distinctive enough to identify an
address uniquely. However, for matching efficiency we prefer distinctive tokens.
We will come back to this topic after we explain how to measure the similarity between two addresses as two bags of tokens.

\section{Similarity between Bags of Tokens}
We assess the similarity between two addresses as the
similarity between two bags of tokens.

We use Jaccard index to measure the similarity between two bags of tokens. 
Jaccard index of two sets is defined as the ratio between the number of common
elements and the number of total elements.
\begin{equation}\label{eq:jac}
J(T_1,T_2)=\frac{|T_1\cap T_2|}{|T_1\cup T_2|}
\end{equation}
For example, consider two bags of tokens 
\begin{eqnarray*}
T_1&=&\{this, is, an, example\}\\
T_2&=&\{this, is, another, example\}\\
T_1\cap T_2&=&\{this, is, example, this, is, example\}\\
T_1\cup T_2&=&\{this, is, an, example, this, is, \\
&&another, example \}\\
J(T_1,T_2)&=&\frac{|T_1\cap T_2|}{|T_1\cup T_2|}=\frac{6}{8}=0.75\quad.
\end{eqnarray*}
As one can see, the Jaccard index between two sets is always in the range
between 0 and 1. Here 0 indicates that two sets have nothing in common, and 1
that the two sets are exactly the same. The more common elements two sets share relative
to the total number of tokens they have, the larger their Jaccard index is. We
say two addresses are equivalent if their Jaccard index exceeds a
threshold $\tau$.

We shall see in Section \ref{sec:discussion} that the algorithm admits other similarity functions too.

\section{Inverted Index}

Na{\"i}vely, linking an address to a database requires comparing this particular
address against each database address to obtain their similarity. Indexed tokens
allow us to do the linking in sublinear time.

We build an inverted index for addresses in the database. An inverted
index keeps all the distinct tokens in the database. For each distinct token,
the inverted index also keeps references to all the addresses which contain this
token.

When a query address arrives, an inverted index allows us to know which database
addresses share common tokens with the query address without scanning through
the database. More specifically, given a query address, we first break this
query address into a bag of tokens $Q$. If a token is not included in the inverted index, we
simply ignore the token. Each remaining token selects a segment
from the inverted index. Database addresses appearing on these segments share at
least one common token with the query address. We can then count the number of
occurrences of each database address $C_i$ on these segments, which gives us the value
of $|Q\cap C_i|$ for each $i$. We then derive the value of $|Q \cup C_i|$ for each $i$ using 
\begin{equation}
|Q \cup C_i|=|Q|+|C_i|-|Q\cap C_i| \quad. 
\end{equation}
We can then calculate the Jaccard index between the query address $Q$ and each candidate address $C_i$ using Eq~\ref{eq:jac}.

With an inverted index, we only compute the Jaccard index between a query address
and those database addresses whose Jaccard indexes are non-zero. The efficiency
of address linkage therefore depends on the number of addresses that share at
least one token with the query address, not the size of the database. 

\section{Two-Round Linkage}

Recall our earlier discussion that tokens of different types have different
distinctiveness. The number of database addresses that contain a more
distinctive token is by definition smaller than the number of database addresses
that contain a less distinctive token. We therefore have better linking
efficiency with more distinctive tokens. Yet in return, we may miss more
matches due to data quality issues.

To maximise linking efficiency while minimising the number of missed matches, we
propose a two-round linkage schema. In the first round, we use distinctive
tokens, \textit{e.g.}, phrase tokens, and inverted indexes to shortlist database
addresses which have non-zero Jaccard indexes with the query address. In the
second round, we compute the Jaccard index between the query and shortlisted
addresses using less distinctive tokens to account for data quality issues.

In this way, the distinctive tokens decide which database entries get involved
in the linkage. The less distinctive tokens decide the similarity between a
query and a database entry. A database entry gets involved as long as it shares
a distinctive token with the query. A database entry matches a query if they
have enough less-distinctive tokens in common.

The two-round linkage strategy is similar to the one described in \cite{Arasu:2006:EES:1182635.1164206}.

\section{A Batch Linkage Algorithm}

Quite often, we need to find equivalent addresses between two large databases
each containing millions of addresses. Na{\"i}vely, we could perform pairwise matching for every combination of addresses.
We describe in this section a simple, and possibly novel, extension of the inverted index data structure to allow efficient linking of two large address databases.

To do batch linking between two databases, we build separate inverted indexes
for each database. From each inverted index, we eliminate all the tokens that
recur more than $k$ times. (More on that soon.) We then join the two inverted indexes by the
common tokens they share. Joining a pair of
common tokens essentially joins two sets of addresses from two databases,
respectively. Every pair of addresses from these two sets is a potential
match. Between these pairs, we then compute the Jaccard
index to identify true matches.

We eliminate tokens that recur more than $k$ times. If a token is
too common, addresses linked by this token are not likely to be a true match.
Moreover, examining addresses linked by a common token takes a lot of time, but
does not find proportionally more matches. Ignoring these common tokens will not
miss many true matches because these matches are usually also linked by some more
distinctive tokens. 

Linking one address at a time can be seen as a special case of batch linkage,
\textit{i.e.}, one of the databases contains only one address. The advantage of
batch linkage over performing a single linkage many times is that in batch
linkage we join the two inverted indexes only once, instead of many times.

Our batch linkage can be explained in the traditional framework of data linkage,
where joining two inverted indexes implements (data-driven) blocking.
Nevertheless, there are also some notable differences.
Instead of using fixed blocking keys like postcode and suburb, we use tokens as blocking keys.
Importantly, deciding which token is used as a blocking key is determined by the data, more specifically its recurring frequency.
This allows the algorithm to adapt to characteristics of the specific databases to be matched.

The above extension of inverted indexes applies to the first of the two-round linkage schemes described above. The second round of pairwise Jaccard calculations of shortlisted candidate address pairs is done using the algorithm described in the following section.

\subsection*{Computing Jaccard Index in Linear Time}
%

We first sort the the tokens in each set. This can be done efficiently since the number of distinct tokens is small. 
We then sort the tokens, and read from the two sets at the same time
following the rules below:
\begin{enumerate}
  \item If the two tokens read in are the same, we increase the number of
  common tokens and the number of total tokens both by 2. We read one more
  token from each set.
  \item If one token is larger than the other, we increase the number of
  total tokens by 1. We read one more token from the set whose
  current token is smaller.
\end{enumerate}
We finish reading when either set is exhausted, and add the number of remaining
tokens in the other set into the number of total tokens. The division
between the number of common tokens and the number of total tokens then provides the
Jaccard index.

For small tokens (like characters or 2-grams), the time complexity of the algorithm is
$O(l+r)$, where $l$ and $r$ denote the number of tokens in the two sets.

\subsection*{SQL Implementation}
The full algorithm in (almost ANSI) SQL is listed in Algorithm~\ref{alg:batch linkage}.
The SQL code runs on Greenplum and PostgreSQL.
The {\tt DISTRIBUTED BY} keyword in table creation specifies how the rows of a table are stored distributively across a cluster by hashing on the distribution key.
The Greenplum database query optimiser will exploit the structure of the SQL query and the underlying data distribution to construct optimal execution plans.

With minor modifications, the SQL code can be modified to run on other parallel databases like Teradata and Netezza, and parallel platforms like Spark (using Spark SQL) and Hadoop (using HIVE, HAWQ \cite{Chang:2014:HMP:2588555.2595636} or Impala \cite{Kornacker2015ImpalaAM}.
It's also straightforward to implement the algorithm in Scala/Python running natively on Spark.

\begin{algorithm}[h!]\small
\caption{SQL Code for Batch Linkage}
\begin{algorithmic}[1]
  \STATE
  CREATE TABLE address\_db \quad \%\% Original address data\\ 
(\quad address\_id bigint,\\
~\quad address text\quad\quad\quad~)\\
DISTRIBUTED BY (address\_id); 
\STATE 
CREATE TABLE address\_db\_phrase\quad  \%\% Compute 2-word phrase tokens\\
(\quad address\_id bigint,\\
~\quad token\_phrase text\quad)\\ 
DISTRIBUTED BY (token\_phrase); 
\STATE INSERT INTO address\_db\_phrase\\
SELECT
address\_id,(regexp\_matches( regexp\_replace(address,'[$^\wedge$A-Z0-9]+',' ','g')\\
\quad ,'[A-Z0-9+]+ [A-Z0-9+]+','g'))[1] \\
FROM address\_db\\
UNION\\
SELECT address\_id,(regexp\_matches( regexp\_replace( regexp\_replace(address,\\
\quad '[$^\wedge$A-Z0-9]',' ','g') ,'[A-Z0-9]+','') ,'[A-Z0-9+]+ [A-Z0-9+]+','g'))[1] \\
FROM address\_db;
\STATE
CREATE TABLE address\_db\_phrase\_inverted\quad  \%\% Compute inverted index \\
(\quad token\_phrase text,\\
~\quad address\_ids bigint[],\\
~\quad frequency bigint\quad~)\\
DISTRIBUTED BY (token\_phrase); 
\STATE INSERT INTO address\_db\_phrase\_inverted\\
SELECT token\_phrase,array\_agg(address\_id),count(1)\\
FROM address\_db\_phrase\\
GROUP BY token\_phrase; 
\STATE
CREATE TABLE address\_db\_phrase\_matched \%\% Matched address arrays \\
(\quad token\_phrase text, \\
~\quad address\_ids\_1 bigint[],\\
~\quad address\_ids\_2 bigint[]\quad)\\
DISTRIBUTED BY (token\_phrase); 
\STATE \%\% {\tt address\_db\_phrase\_inverted\_2} is the second dataset.\\
  INSERT INTO address\_db\_phrase\_matched\\
SELECT l.token\_phrase,l.address\_ids,r.address\_ids\\
FROM address\_db\_phrase\_inverted\_1 AS l\\
INNER JOIN address\_db\_phrase\_inverted\_2 AS r\\
ON l.token\_phrase=r.token\_phrase 
AND l.frequency$\leq100$ AND r.frequency$\leq100$; 
\STATE
CREATE TABLE address\_db\_proposed\_match  \%\% Unnest candidate address pairs \\
(\quad address\_id\_1 bigint,\\
~\quad address\_id\_2 bigint ~)\\
DISTRIBUTED BY (address\_id\_1); 
\STATE INSERT INTO address\_db\_proposed\_match\\ 
SELECT DISTINCT address\_id\_1, unnest(address\_ids\_2)\\
FROM ( SELECT unnest(address\_ids\_1) AS address\_id\_1, address\_ids\_2 \\
\quad\quad\quad\quad FROM address\_db\_phrase\_matched ) AS tmp;
\STATE
CREATE TABLE address\_db\_match AS \quad  \%\% Compute round 2 Jaccard index\\
SELECT address\_id\_1, address\_id\_2, jaccard(t2.address, t3.address) \\
FROM address\_db\_proposed\_match t1, \\
\quad address\_db\_1 t2, \\
\quad address\_db\_2 t3 \\
WHERE t1.address\_id\_1 = t2.address\_id \\
AND t1.address\_id\_2 = t3.address\_id 
\end{algorithmic}
\label{alg:batch linkage}
\end{algorithm}

\section{Experiments}

We demonstrate the performance of our proposed algorithm in two scenarios: linking an address dataset against a reference address dataset, and linking two arbitrary address datasets.
In the first scenario, for each address in the first dataset, it can be assumed that there exists a match in the reference dataset.
In the latter scenario, we have to provide for the case where there is no match for an address.

\subsection{Linking with a Reference Dataset}
This scenario usually occur during address cleansing. We deal with two address
databases. The first database contains raw addresses, whereas the second database contains reference
addresses. For each raw address, we search for its equivalent
reference address, which provides a cleansed
representation of the raw address.

In this experiment, we use two address databases:
\begin{itemize}
  \item \textbf{AGA1} is a raw database collected by an Australian Government Agency.
    The database contains around 48 millions
  addresses most of which are Australian addresses. Addresses in this database
  are known to have significant data quality issues, with many incomplete and inaccurate addresses.

\item \textbf{OpenAddress\_Australia} contains more than 19
  millions Australian addresses. All addresses are in standard form. This
  reference address database is open-source and can be downloaded from \url{https://openaddresses.io}.
  Almost all Australian addresses in AGA1 
  have a reference entry in OpenAddress\_Australia.
\end{itemize}

We use the batch linkage algorithm to link addresses in AGA1 with addresses in
OpenAddress\_Australia. We extract order-preserving 2-word phrase tokens from the addresses and
construct inverted indexes for both databases. We then compute character-based
Jaccard index between each pair of shortlisted candidates. We accept a link if
the Jaccard index exeeds a threshold $\tau$.

Since we do not have a ground truth for the address cleansing
result, we can not quantitatively assess the rate of false negatives (i.e.
there exists a cleansed entry for a raw address but the algorithm cannot find
it) in our linkage result. It is fair to say that essentially all data operations
involving large databases have the same problem. It is therefore difficult to
select the proper threshold value $\tau$. We propose the following mechanism for threshold
selection. We implement address linkage with increasing thresholds,
\textit{e.g.}, $\{\tau_1=0.6,\tau_2=0.7,\tau_3=0.8\}$. We then use the result
of the lowest threshold to benchmark that of higher thresholds for false
negatives.

\begin{figure}[h]
\centering
\includegraphics[width=0.8\textwidth]{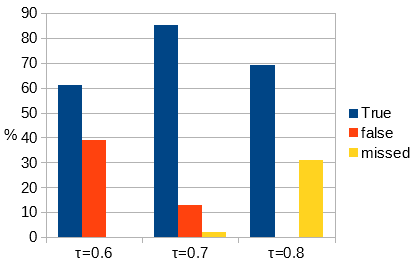}
\caption{\label{fig:histo} the percentage of true positives (true), false
positives (false), and false negatives (missed) of proposed addressing linkage
algorithm with different thresholds.}
\end{figure}

Figure~\ref{fig:histo} shows the percentage of true positives, false positives,
and false negatives for the proposed method. These results are obtained by
manually assessing 100 randomly sampled linked addresses. As we can see, when
$\tau=0.6$, which roughly requires a cleansed address to share $60\%$ or more
characters with the raw address, nearly $40\%$ of raw addresses will find false
cleansed forms. When $\tau$ increases to $0.7$, the percentage of false
positives drops to $12\%$. Conversely, $2\%$ of raw addresses
which used to find cleansed forms can no longer find them. This missing rate
rises to $31\%$ when $\tau$ increases to $0.8$. Among the three values,
$\tau=0.7$ gives the best performance.

Table~\ref{tb:open} lists 10 example links between AGA1
and OpenAddress\_Australia found by our algorithm. Due to privacy concern, the addresses in these examples have been modified and does not reflect the original addresses.
Besides the two linked addresses, we also provide the clean addresses found by Google Maps for AGA1
addresses. To protect anonymity we have encrypted the street
names and suburb names. Interestingly, there are three
addresses that Google Maps failed to process, yet were successfully linked by our algorithm.
The fourth example in the table shows the limitation of using characters as tokens in the Jaccard index calculation.
A simple tie-breaker postprocessing scheme using, for example edit distance, can be used to resolve such issues.

\begin{table}[ht!]\centering
\caption{\label{tb:open}Address Linkage between AGA1 and OpenAddress\_Australia}
\begin{small}
\begin{tabular}{llc}

\hline
&Address& Jaccard\\
\hline
AGA3&33 34-38 EHMNTV DIU NSW 6561\\
Open&UNIT 33 34-38 EHMNTV STRET OUT DIR NSW 6561&0.88\\
Google&33/34-38 EHMNTV ST OUT DIU NSW 6561\\
\hline
AGA3&53 741 ADGNR EFORST AKLR QLD 9368\\
Open&UNIT 53 741 ADGNR AVENUE EFORST LAKE QLD 9168&0.80\\
Google&53/741 ADGNR AVE EFORST AEKL QLD 9168\\
\hline
AGA3&972 4 CEOPRW LMOW NEW WALES 5133\\
Open&UNIT 972 4 CEOPRW AFHRW ROADWAY LMOW NSW 5133&0.90\\
Google&NOT FOUND\\
\hline
AGA3&713 311 GUN HILNU ACT 5035\\
Open&UNIT 731 311 GUN PLACE HILNU ACT 5035&0.93\\
Open&UNIT 713 311 GUN PLACE HILNU ACT 5035&0.93\\
Open&UNIT 317 311 GUN PLACE HILNU ACT 5035&0.93\\
Google&713/311 GUN PL HILNU ACT 5035\\
\hline
AGA3& 3 59 FGIS DEKNOR QLD QLD 9173\\
Open& UNIT 3 59 FGIS STRET DEKNOR QLD 9173&0.90\\
Google&3/59 FGIS ST DEKNOR QLD 9173\\
\hline
AGA3&9 NO 7 TO 2 CELMNT ADEGNO VIC 7362\\
Open&UNIT 9 7-2 CELMNT STRET ADEGNO VIC 7362&0.91\\
Google& NOT FOUND\\
\hline
AGA3&313 0 EGKNORW ABEHILTZ BAY 5133\\
Open&UNIT 313 0 EGKNORW AVENUE ABEHILTZ BAY NSW 5133&0.91\\
Google&313/0 EGKNORW AVE ABEHILTZ BAY NSW 5133\\
\hline
AGA3&MARGETIC 6 715 ABDFORST BELMNORU VIC 7123\\
Open&FLAT 6 715 ABDFORST STRET HNORT BELMNORU VIC 7123&0.92\\
Google&NOT FOUND\\
\hline
AGA3&43 3345 ACDHINSV AGRTV QLD 9355\\
Open&UNIT 43 3345 ACDEHINSV ROAD MOUNT AGRTV EAST QLD 9355&0.82\\
Google&43/3345 ACDEHINSV RD MOUNT AGRTV EAST QLD 9355\\
\hline
AGA3&78 03 ADELMNOR BELM VIC VIC 7133\\
Open&UNIT 78 03 ADELMNOR STRET ACFORSTY VIC 7133&0.83\\
Google&78/03 ADELMNOR ST ACFORSTY VIC 7133\\
\hline    
\end{tabular}
\end{small}
\end{table}

\subsection{Linking Two Arbitrary Datasets}
This scenario occurs when people try to integrate two databases together.
%
To test this scenario, we use two databases AGA1 and AGA2.
\begin{itemize}
  \item \textbf{AGA2} contains around 18 millions addresses
  collected by a large Australian government department. Most addresses in AGA2 are Australian
  addresses. Addresses in this database may be incomplete and inaccurate. AGA1
  and AGA2 are collected by different government agencies from different sources and for largely different original purposes.
   
\end{itemize}

We again use the batch linkage algorithm with 2-word phrase tokens for round 1 of Jaccard computations and character tokens for round 2.
However, in this second address-linkage scenario, we can no longer use a simple threshold $\tau$ to reject
false matches. This is because when linking with a reference dataset, if a
street is included in the reference database, all individual addresses
in this street are included. Therefore, if a raw address has a high score best
match in the reference database, this best match is usually consistent with the
raw address in every detail.
However, in the scenario where we are linking two arbitrary databases, it is quite common for two databases to contain only two different addresses in the same street. These two addresses may have the highest matching score but still remain a false match.
To complicate matters, a true match can also be a low score match due
to data quality issues with both addresses.

One way to overcome this challenge is to require two matching addresses to have
consistent numeric tokens. We say two sets of numeric tokens are
consistent, if one set is a subset of the other.

\begin{figure}[h]
\centering
\includegraphics[width=0.8\textwidth]{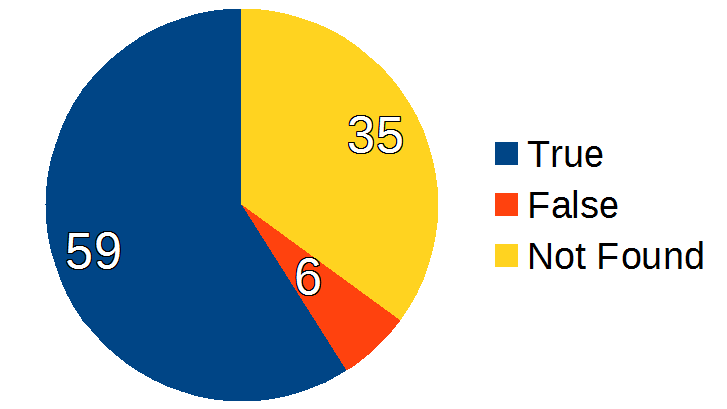}
\caption{\label{fig:pie}Percentage of correctly linked (True), incorrectly
linked (False), and not linked (Not Found) when joining AGA2 addresses to AGA1 addressees using proposed algorithm.}
\end{figure}

We manually assess 100 randomly sampled AGA2 addresses. For each AGA2 address,
we in order consider its top 3 matches in AGA1 database. If a match has
consistent numeric tokens and is a true match, we label this AGA2 sample as true
and no longer consider the remaining matches. If a match has consistent numeric tokens
but is a false match, we label this AGA2 sample as false and no longer consider
the remaining matches. If none of the top 3 matches has consistent numeric tokens with
the query, this AGA2 sample is labelled as not found. Figure~\ref{fig:pie}
shows the percentage of three labels in the 100 samples. It can be derived from
Figure~\ref{fig:pie} that, $59/(59+6)=91\%$ of the samples are correctly linked.

\begin{table*}[ht!]
\centering
\caption{\label{tb:link}Address Linkage between AGA1 and AGA2}
\begin{small}
\begin{tabular}{clc}
\hline
&Address& Jaccard\\
\hline
AGA1&4 EGNORU AEHLMT VIC 3095\\
AGA2&4 EGNORU CRT AEHLMT NORTH VIC 3095&0.84\\
\hline
AGA1&528 LTUY HLU LTUY HLU QLD 4854\\
AGA2&528 LTUY ADEHLSU RD LTUY QLD 4854&0.87\\
\hline
AGA1&45 EGHIMNS ADEGNO VIC 3175\\
AGA2&RM 8 45 EGHIMNS ST ADEGNO VIC 3175&0.88\\
\hline
AGA1&6 EILS CEIMNRTY ABILRSUY DNOSW SA 5108\\
AGA2&6 EILS CEIMNRTY RD ABILRSUY DNOSW SA 5108&0.96\\
\hline
AGA1&EL EILOSU 137 AILNT EFNRY EOV QUEN SLAND 4055\\
AGA2&137 AILNT RD EFNRY EGORV QLD 4055&0.74\\
\hline
AGA1&80 ABEGLNRU DEHILS VI 3037\\
AGA2&80 ABEGLNRU DR DEHILS VIC 3037&0.94\\
\hline
AGA1&141 ACEHLRS EHPRT 6005\\
AGA2&141 ACEHLRS ST ESTW EHPRT WA 6005&0.80\\
\hline
AGA1&51 BENOR BELMNOT VIC 3216\\
AGA2&2/51 BENOR DR BELMNOT VIC 3216&0.91\\
\hline
AGA1&97 ELOXY ABDPRUY WA 6025 TRA LIA\\
AGA2&97 ELOXY AVE ABDPRUY WA 6025&0.87\\
\hline
AGA1&9 DLORS DENSY 2077\\
AGA2&9 DLORS AVE AHIQSTU NSW 2077&0.68\\
\hline    
\end{tabular}
\end{small}
\end{table*}

Table~\ref{tb:link} lists 10 example links between AGA1 and AGA2 found by our algorithm.

\subsection{Computational Efficiency}
When dealing with a large database, algorithm efficiency is as important as
algorithm accuracy, because an algorithm that takes days to finish
is too expensive to deploy, and even more expensive to test under multiple
configurations. Experiments show that our algorithm is highly efficient and scalable
to large databases.

Using the open-source Greenplum Database running on 8 servers (1 master + 7 slaves), each with 20 cores, 320 GB, and 4.5 TB usable RAID10 space, linking 48 million AGA1 addresses with 13 million OpenAddress addresses using our algorithm takes about 5 minutes.
Linking 48 million AGA1 addresses with 18 million AGA2 addresses takes about 7.5 minutes.
The algorithm also scales essentially linearly in the number of servers in the Greenplum cluster dedicated to the task.

Note that the processing time of our algorithm depends more on the
similarity between two databases than on the sizes of the two databases.
The efficiency of the algorithm is due to the following factors:
\begin{enumerate}
   \item The quantity of Jaccard index computation does not depend on the size of
   the database, but the number of addresses sharing common distinctive tokens.
   \item Finding addresses sharing common distinctive tokens is done jointly for
   all addresses at the same time. This overhead does not depend on the number of
   addresses, but the number of distinctive tokens during the joining between two
   inverted indexes.
\end{enumerate}

\section{Parameters of the Algorithm}
The use of Jaccard index to assess similarity between addresses in our algorithm is optional.
Our implicit assumption is that there exists a function $d(x,y)$ which assesses
the similarity between two addresses $x$ and $y$. Blocking can reduce
the number of evaluations of $d(x,y)$ without missing links, if $d(x,y)>\tau$
indicating $x$ and $y$ share a common token. In our two-round linkage, our implicit
function is
\begin{equation}
 d(x,y)=\left\{
 \begin{array}{ll}
 J_{char}(x,y)& \mbox{if } J_{phrase}(x,y)>0\\
 0 & \mbox{otherwise}
 \end{array}
 \right.
\end{equation}
One may design any other implicit function instead, replacing the Jaccard index with any other measurement.

The Jaccard index in round 2 of the comparison can also be replaced by almost any other similarity function, for example the Monge-Elkan function \cite{Monge97anefficient}, which is suitable for addresses.



\section{Conclusion}\label{sec:discussion}

We have presented in this paper a novel address-linkage algorithm that:
\begin{enumerate}\itemsep1mm\parskip0mm
\item links addresses as free text;
\item uses data-driven blocking keys;
\item extends the inverted index data structure to facilitate large-scale address linking;
\item is robust against data-quality issues; and
\item is practical and scalable. 
\end{enumerate}



The simplicity of the solution - a great virtue in large-scale industrial applications - may belie the slightly tortuous journey leading to its discovery; a journey laden with the corpses of a wide-range of seemingly good ideas like compressive sensing and other matrix factorisation and dimensionality-reduction techniques, nearest-neighbour algorithms like KD-trees, ElasticSearch with custom rescoring functions \cite{elasticsearch}, rules-based expert systems, and implementation languages that range from low-level C, to R, Python, SQL and more.
In retrospect, our algorithm can be interpreted as an application of a signature-based approach to efficiently compute set-similarity joins \cite{Arasu:2006:EES:1182635.1164206}, where the abstract concept of sets is replaced with carefully considered set-representations of addresses, with a modern twist in its implementation on state-of-the-art parallel databases to lift the algorithm's scalability to potentially petabyte-sized datasets.

\bibliographystyle{splncs03}
\bibliography{address_sig}

\begin{thebibliography}{10}
\providecommand{\url}[1]{\texttt{#1}}
\providecommand{\urlprefix}{URL }

\bibitem{Arasu:2006:EES:1182635.1164206}
Arasu, A., Ganti, V., Kaushik, R.: Efficient exact set-similarity joins. In:
  Proceedings of the 32nd International Conference on Very Large Data Bases.
  pp. 918--929. VLDB Endowment (2006)

\bibitem{CPA:CPA20124}
Candès, E.J., Romberg, J.K., Tao, T.: Stable signal recovery from incomplete
  and inaccurate measurements. Communications on Pure and Applied Mathematics
  59(8),  1207--1223 (2006)

\bibitem{Chang:2014:HMP:2588555.2595636}
Chang, L., Wang, Z., Ma, T., Jian, L., Ma, L., Goldshuv, A., Lonergan, L.,
  Cohen, J., Welton, C., Sherry, G., Bhandarkar, M.: {HAWQ}: A massively
  parallel processing {SQL} engine in {H}adoop. In: Proceedings of the 2014 ACM
  SIGMOD International Conference on Management of Data. pp. 1223--1234. ACM,
  New York, NY, USA (2014)

\bibitem{Christen:2012:DMC:2344108}
Christen, P.: Data Matching: Concepts and Techniques for Record Linkage, Entity
  Resolution, and Duplicate Detection. Springer (2012)

\bibitem{Christen05automatedprobabilistic}
Christen, P., Belacic, D.: Automated probabilistic address standardisation and
  verification. In: Australasian Data Mining Conference (AusDM05) (2005)

\bibitem{Christen2004}
Christen, P., Churches, T., Hegland, M.: Febrl -- a parallel open source data
  linkage system. In: Proceedings of PAKDD 2004. pp. 638--647. Springer (2004)

\bibitem{stringmetrics}
Cohen, W.W., Ravikumar, P., Fienberg, S.E.: A comparison of string metrics for
  matching names and records. In: SIGKDD (2003)

\bibitem{elasticsearch}
Gormley, C., Tong, Z.: Elasticsearch: The Definitive Guide. O'Reilly Media
  (2015)

\bibitem{Guo:2009:ASL:1557019.1557144}
Guo, H., Zhu, H., Guo, Z., Zhang, X., Su, Z.: Address standardization with
  latent semantic association. In: Proceedings of the 15th ACM SIGKDD
  International Conference on Knowledge Discovery and Data Mining. pp.
  1155--1164. KDD '09, ACM, New York, NY, USA (2009)

\bibitem{halevy09}
Halevy, A., Norvig, P., Pereira, F.: The unreasonable effectiveness of data.
  IEEE Intelligent Systems  24,  8--12 (2009)

\bibitem{DBLP:journals/pvldb/HellersteinRSWFGNWFLK12}
Hellerstein, J.M., R{\'{e}}, C., Schoppmann, F., Wang, D.Z., Fratkin, E.,
  Gorajek, A., Ng, K.S., Welton, C., Feng, X., Li, K., Kumar, A.: The {MAD}lib
  analytics library or {MAD} skills, the {SQL}. {PVLDB}  5(12),  1700--1711
  (2012)

\bibitem{Kornacker2015ImpalaAM}
Kornacker, M., Behm, A., Bittorf, V., Bobrovytsky, T., Ching, C., Choi, A.,
  Erickson, J., Grund, M., Hecht, D., Jacobs, M., Joshi, I., Kuff, L., Kumar,
  D., Leblang, A., Li, N., Pandis, I., Robinson, H., Rorke, D., Rus, S.,
  Russell, J., Tsirogiannis, D., Wanderman-Milne, S., Yoder, M.: Impala: A
  modern, open-source {SQL} engine for {H}adoop. In: CIDR (2015)

\bibitem{Monge97anefficient}
Monge, A., Elkan, C.: An efficient domain-independent algorithm for detecting
  approximately duplicate database records (1997)

\bibitem{Zaharia:2010:SCC:1863103.1863113}
Zaharia, M., Chowdhury, M., Franklin, M.J., Shenker, S., Stoica, I.: Spark:
  Cluster computing with working sets. In: Proceedings of the 2Nd USENIX
  Conference on Hot Topics in Cloud Computing. pp. 10--10. HotCloud'10, USENIX
  Association, Berkeley, CA, USA (2010)

\end{thebibliography}
\end{document}